\documentclass[final,5p,times,twocolumn]{elsarticle}

\usepackage{graphicx}
\usepackage{amssymb}
\usepackage{lineno}

\journal{Nuclear Instruments and Methods A}

\begin{document}

\begin{frontmatter}



\title{Particle Showers in a Highly Granular Hadron Calorimeter}

% if there is only one institution, use this form:
%\author{John Author, Giovanna Autore}
%\address{University of Wisdom, Physics City, Scienceland}

% else, use optional labels to link authors explicitly to addresses,
% as shown below:
\author[A,B]{Katja Seidel}
\author{for the CALICE collaboration}
\address[A]{Max-Planck-Institute f\"ur Physik, Munich, Germany}
\address[B]{Excellence Cluster 'Universe', TU M\"unchen, Garching, Germany}
\begin{abstract}
The CALICE collaboration has constructed highly granular electromagnetic and hadronic
calorimeter prototypes to evaluate technologies for the use in detector systems at a
future Linear Collider. The hadron calorimeter uses small scintillator cells
individually read out with silicon photomultipliers. The system with 7608 channels
has been successfully operated in beam tests at DESY, CERN and Fermilab since 2006,
and represents the first large scale tests of these devices in high energy physics
experiments. The unprecedented granularity of the detector provides detailed
information of the properties of hadronic showers, which helps to constrain hadronic
shower models through comparisons with model calculations. Results on longitudinal 
and lateral shower profiles, compared to a variety of hadronic 
shower models, first results with a software compensation technique for the energy 
resolution and an outlook on the next generation detector prototype are presented.

\end{abstract}


\end{frontmatter}


\section{Introduction}
The CALICE collaboration \cite{calice} has constructed highly granular calorimeter prototypes to evaluate technologies for the use in detector systems at a future Linear Collider \cite{ilc}. Calorimeter physics prototypes were tested in various different configurations in particle beams at DESY, CERN and FNAL. The presented results were obtained with pion beams at CERN in 2007 and a calorimeter setup consisting of a silicon-tungsten electromagnetic calorimeter (ECAL) \cite{ecal}, an analog scintillator-steel hadron calorimeter (AHCAL) \cite{ahcal} and a scintillator-steel tail catcher and muon tracker (TCMT) \cite{tcmt}.\\
The ECAL has a total depth of 24 $X_0$ and consits of 30 layers with three different sampling fractions. The first 10 layers have 1.4\,mm thick tungsten absorber plates, the next 10 layers have 2.8\,mm thick absorber plates and the last 10 layers have 4.2\,mm thick plates. Each active layer has an area of 18$\times$18\,$\textrm{cm}^2$, which is read out by 1$\times$1\,$\textrm{cm}^2$ readout pads.
\\
The AHCAL consists of 3\,cm thick stainless steel absorber plates and active areas of scintillator tiles with three different sizes, which are read out with silicon photomultipliers (SiPMs). The tiles sizes vary from 3$\times$3\,$\textrm{cm}^2$ in the middle to 12$\times$12\,$\textrm{cm}^2$ at the outer edges. The last eight layers consists only of tiles with sizes of 6$\times$6\,$\textrm{cm}^2$ and 12$\times$12\,$\textrm{cm}^2$. The analog hadron calorimeter has in total 38 layers , which results in a thickness of 4.5 interaction lengths and 7608 scintillator tiles.
\\
The TCMT total thickness corresponds to 5.8 interaction lengths. It consists of 16 layers of 20 100$\times$5\,$\textrm{cm}^2$ scintillator strips, read out by SiPMs, between steel absorber plates. The first 8 absorber layers have a thickness of 19\,mm the last 8 layers have 102\,mm thick plates. The orientation of the scintillator strips alternates between vertical and horizontal in adjacent layers.
\\
In the following, preliminary results of the CALICE collaboration are presented on studies of longitudinal and radial hadronic shower profiles in the AHCAL as well as a software compensation technique to improve the energy resolution of the complete CALICE calorimeter setup. The work on a realistic mechanical structure and compact design of a hadron calorimeter prototype of the ILC is presented in section \ref{sec:next-generation}.

\section{Properties of Hadronic Showers in the AHCAL}
\label{sec:shower-properties}
\begin{figure}[htb!]
 	\centering
 	\includegraphics[trim=0mm 0mm 0mm 160mm, clip, width=0.45\textwidth]{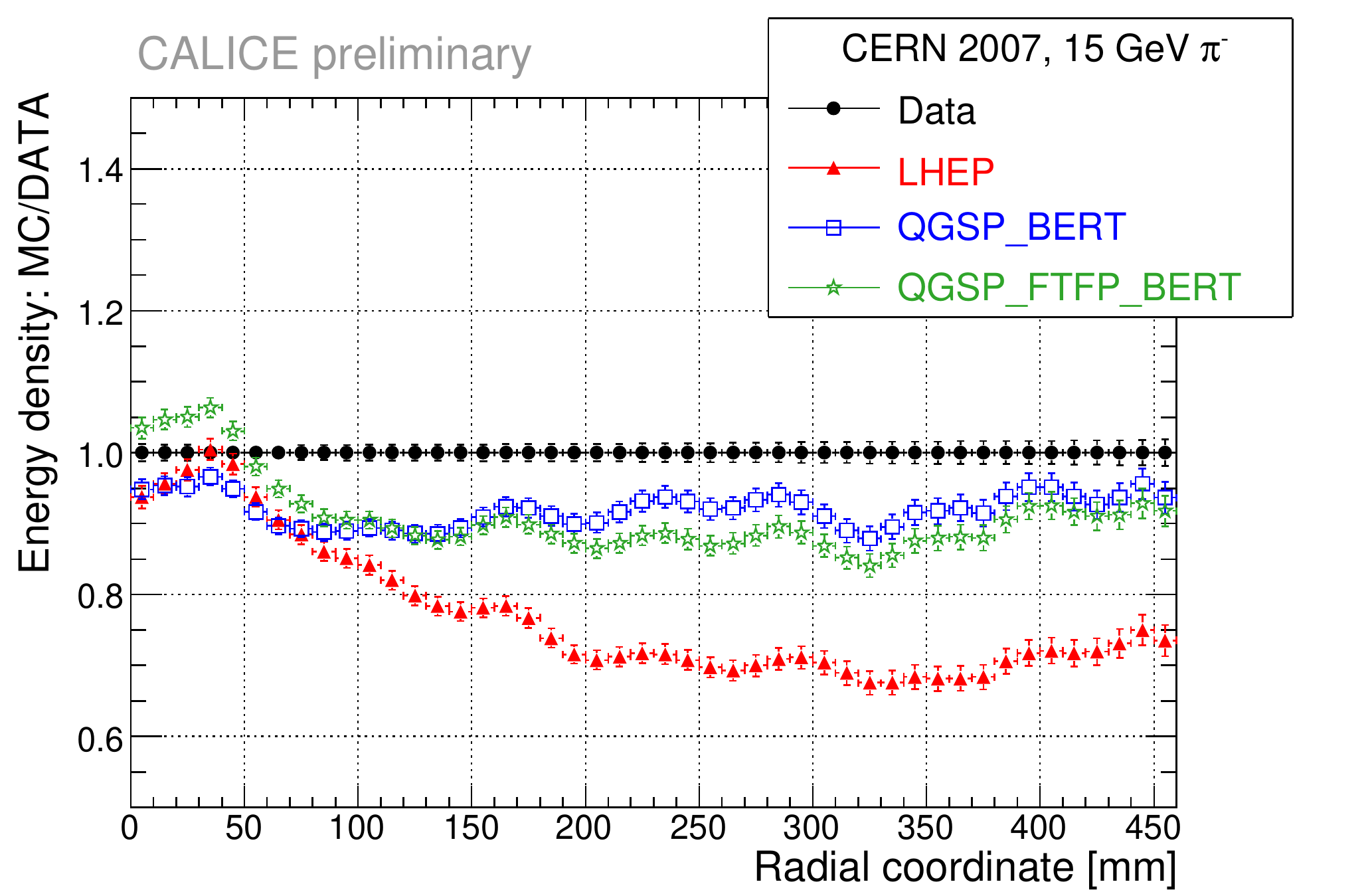}
 	\caption{Relative deviation of the energy from simulations to data for various radial coordinates.}
 	\label{fig:trans-rel}
 \end{figure}

The high granularity allows to compare hadronic shower properties of test beam data and simulations with a high precession. The simulations were produced with different hadronic showers models of GEANT4; the presented results show the models LHEP, QGSP\_BERT, QGSP\_FTF\_BERT, FTF\_BIC, QGSC\_CHIPS and FTFP\_BERT\_TRV \cite{physics-lists}.
\begin{figure}[!htb]
 	\centering
 	\includegraphics[width=0.45\textwidth]{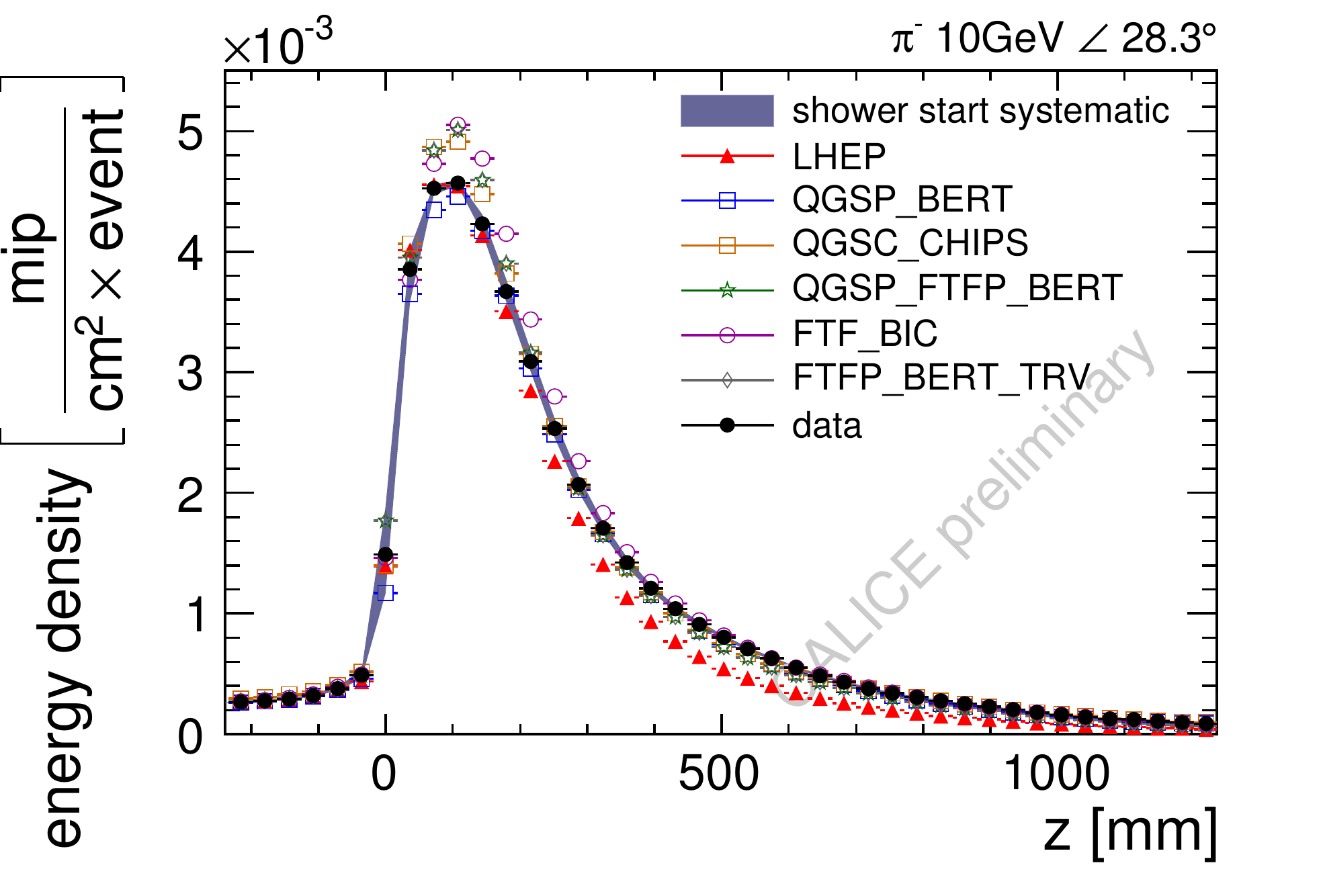}
 	\caption{Longitudinal Profile of 10\,GeV negative pions in a rotated calorimeter for data and simulations.}
 	\label{fig:long-profile}
 \end{figure}
\\The longitudinal profile from the shower start for 10\,GeV negative pions in the AHCAL which was rotated by $28.3^{\circ}$ \cite{Profile} was studied. The shower starting point was found with a detailed analysis on the spatial position of the first hard interaction. If a region of energy deposition exceeds thresholds on the amount of energy deposition and the spatial size and if this region is nearest to the origin of the particle, it was chosen to be the shower start cluster. The near end of the main axis of the shower start cluster defined the shower starting point. 
\\ In Figure \ref{fig:long-profile} the longitudinal profiles for test beam data and simulations are shown and can be compared. LHEP shows the largest deviation from data for higher depths in the calorimeter. FTF\_BIC, QGSC\_CHIPS and FTF\_BERT\_TRV seem to overestimate the energy density at the shower maximum. QGSP\_BERT instead seems to describe the longitudinal profile of the data well over the full extent of the shower. At larger energies, however, discrepancies of data and all considered models have been observed.
\begin{figure}[htb!]
 	\centering
 	\includegraphics[trim=0mm 0mm 0mm 160mm, clip, width=0.45\textwidth]{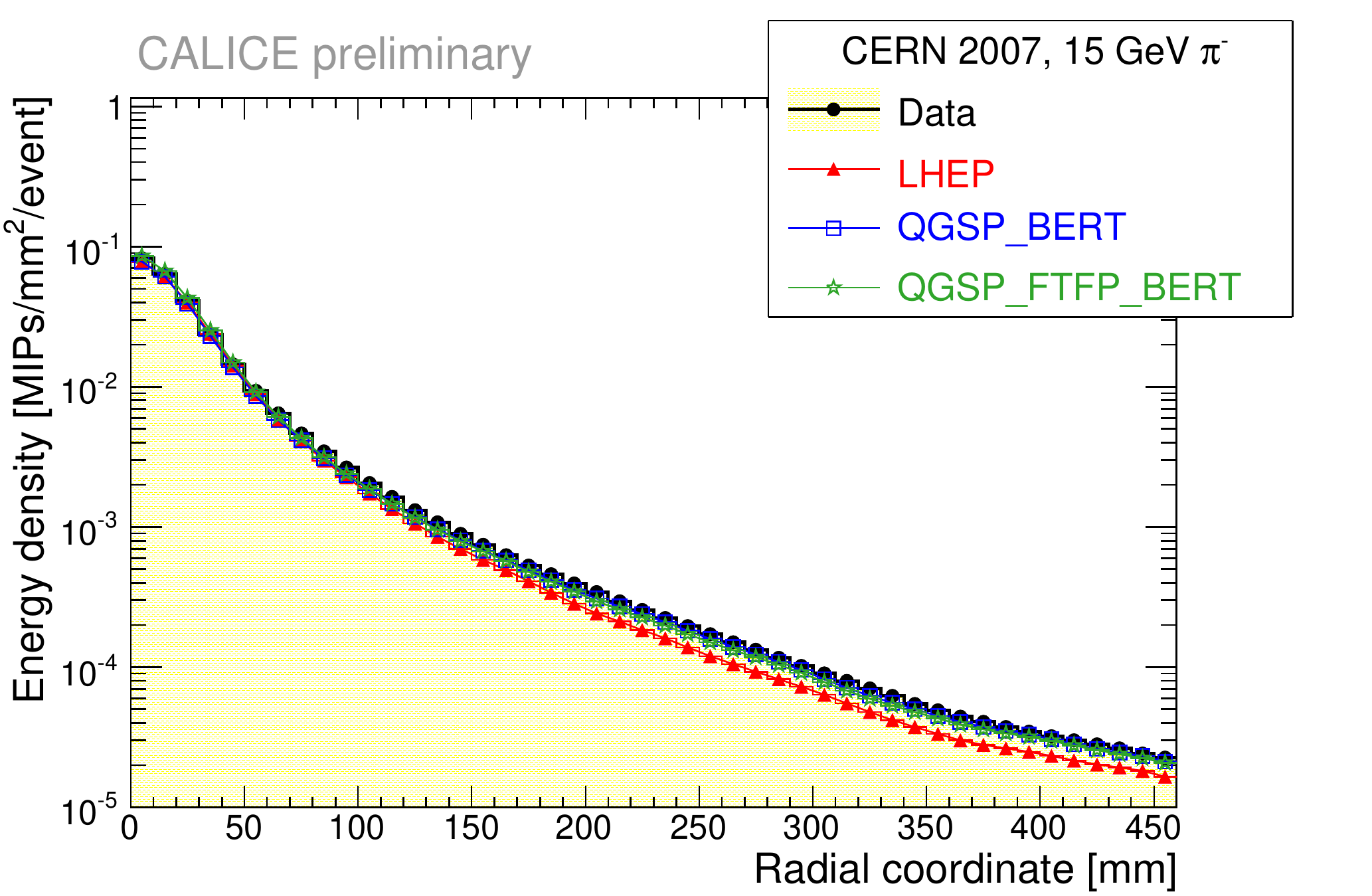}
 	\caption{Transverse shower profile of 15\,GeV pions shower. The energy density is shown as a function of the radial coordinate.}
 	\label{fig:trans-profile}
 \end{figure}
\\A transverse profile of 15\,GeV negative pions is shown in Figure \ref{fig:trans-profile} \cite{Profile}, which shows the energy density for different radial coordinates of the shower. The shower energy density is defined here as the mean energy sum per event per unit of ring area. The ring is concentric around the axis defined by the beam direction and the measured impact point of the particle on the front face of the calorimeter. For a 15\,GeV pion shower the mean radial shower width is approximately 78\,mm. Figure \ref{fig:trans-rel} shows the deviations of the various hadronic shower models compared to the test beam data. Again, LHEP shows the largest  discrepancy to the data. But also the other shower models also give a too small energy density over most of radial coordinate range.

\section{Energy Resolution with Shower Density Dependent Signal Weighting}
\label{sec:energy-resolution}
\begin{figure}[htb!]
 	\centering
 	\includegraphics[width=0.45\textwidth]{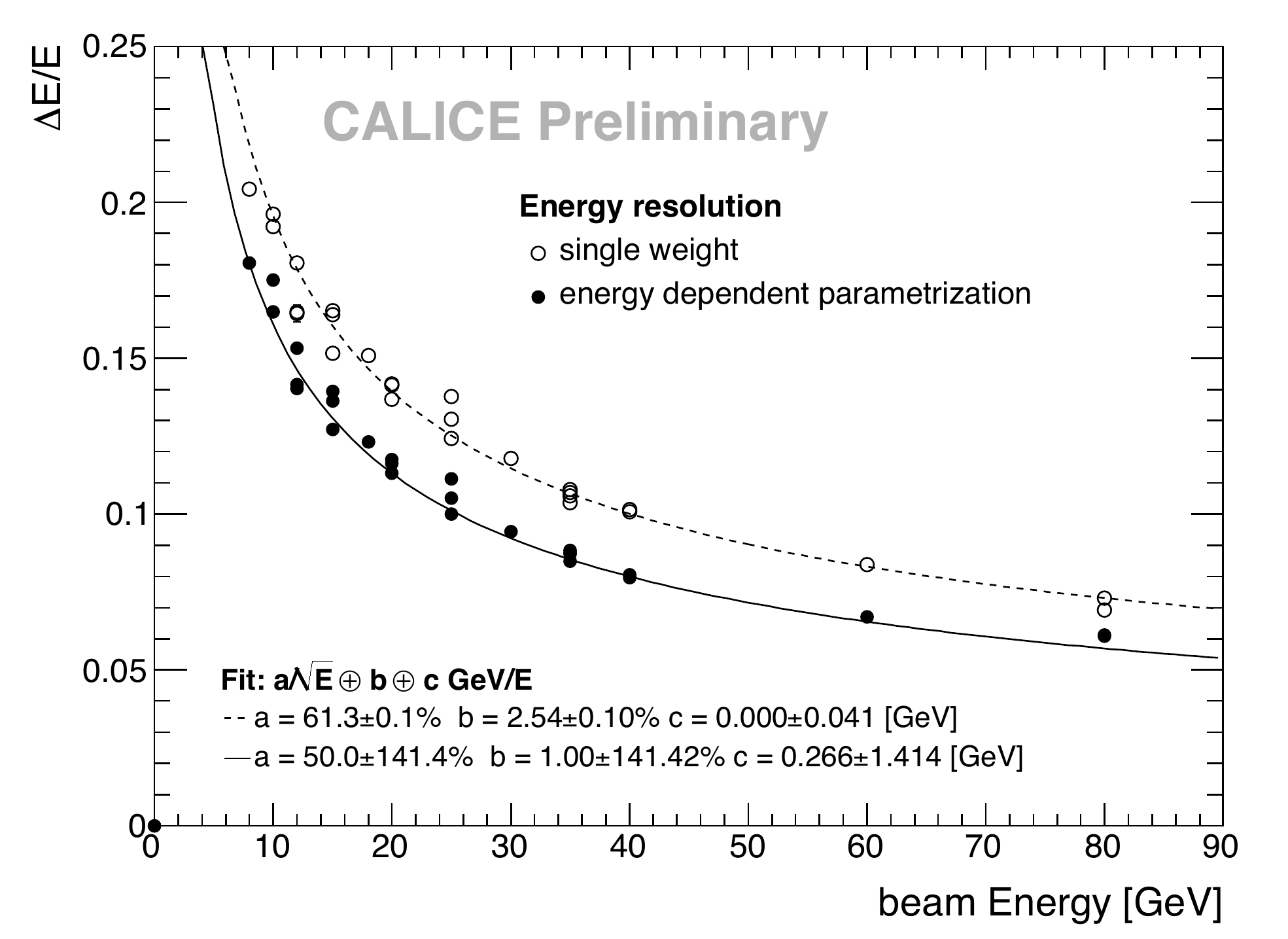}
 	\caption{Energy resolution of the complete CALICE setup for hadrons versus the beam energy. The energy resolution using one factor to calculated the reconstructed energy from the detector signal is shown in the black points and with a local energy density weighting technique applied in the black circles.}
 	\label{fig:resolution}
 \end{figure}

\begin{figure}[htb!]
 	\centering
 	\includegraphics[width=0.45\textwidth]{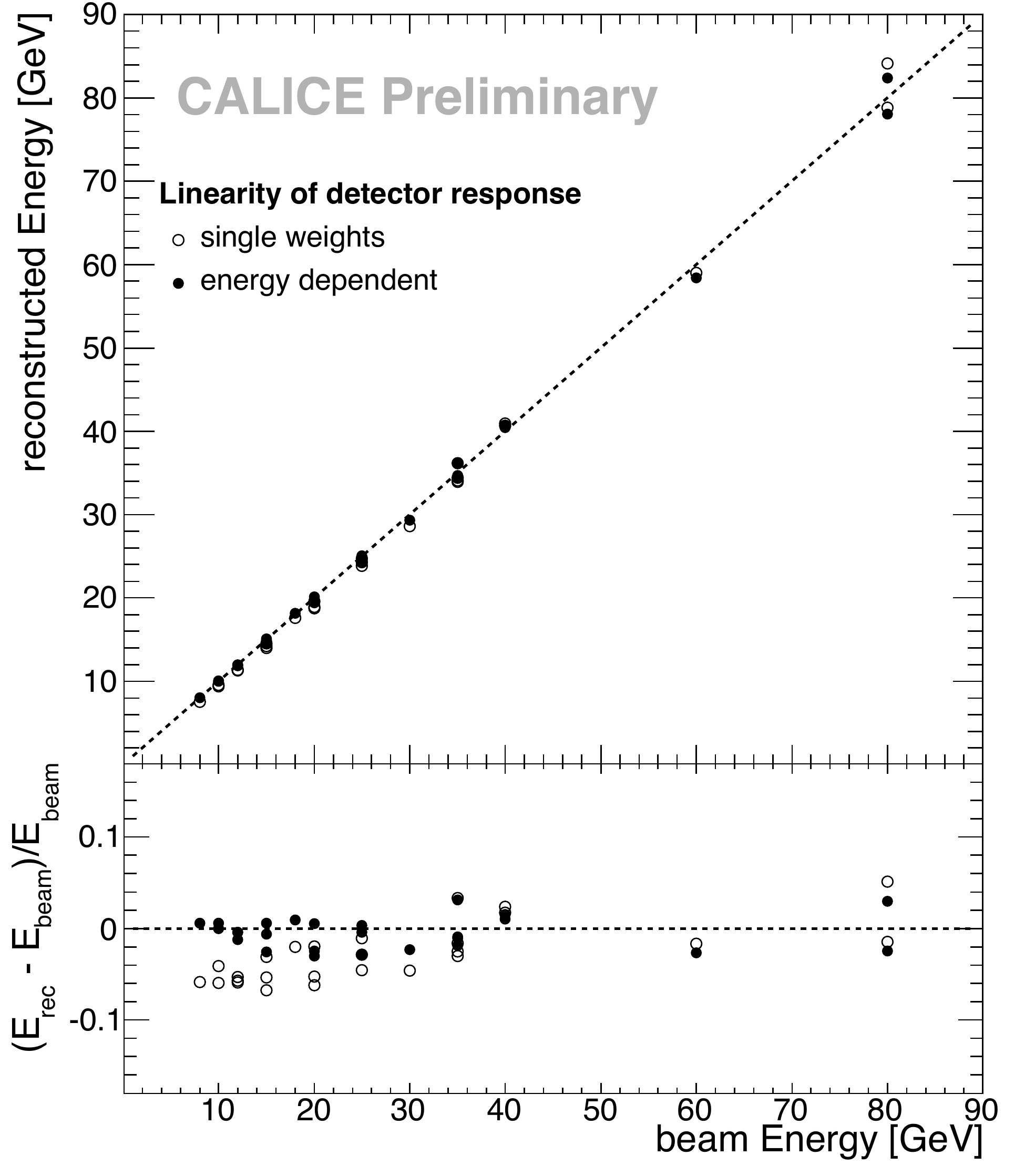}
 	\caption{Reconstructed energy of the complete CALICE setup for hadrons versus the beam energy. The reconstruction obtained with one factor is shown in the black points and with density dependent weighting in the black circles. The lower part shows the relative deviation of the reconstructed from the beam energy.}
 	\label{fig:linearity}
 \end{figure}

The high granularity of the CALICE calorimeters was also used to study software compensation procedures. Hadronic showers, consisting of electromagnetic and hadronic components, show large fluctuations on the fractions of these components from event to event. This leads to non-linearities in the response of the detector and reduces the energy resolution. The energy resolution can be improved if the energy of electromagnetic and hadronic subshowers are treated differently in the overall reconstructed energy, hence if a software compensation is applied.\\
Since electromagnetic showers tend to be denser than hadronic onces, the energy density can be used as a value to discriminate between hits in electromagnetic or hadronic subshowers. In the presented study, hits with a higher energy density are assigned a lower weight in the overall energy sum to compensate for the typically lager detector response to electrons. The weights were determined from statistically independent data sets and an energy dependent parameterization of these weights was found \cite{resolution}. This permits the application of the compensation algorithms without requiring prior knowledge of the beam energy. \\
The initial reconstructed energy was calculated with separate weight factor for ECAL, AHCAL and TCMT. The reconstructed energy and therefore the energy resolution was compared to the one determined with the weighting technique. Figure \ref{fig:resolution} shows the energy resolution of the full CALICE detector setup obtained with and without the weighting technique. An improvement of about 20\,$\%$, to 49\,$\%$ in the stochastic term could be obtained. No requirements on the shower start, the position of the shower or on leakage were made.\\
The linearity, shown in Figure \ref{fig:linearity}, also improved significantly with the weighting technique. The difference between reconstructed and beam energy was obtained to be better than 3\,$\%$.

\section{The Technical Prototype of the Analog Hadron Colarimeter}
\label{sec:next-generation}
\begin{figure}
 	\centering
 	\includegraphics[width=0.45\textwidth]{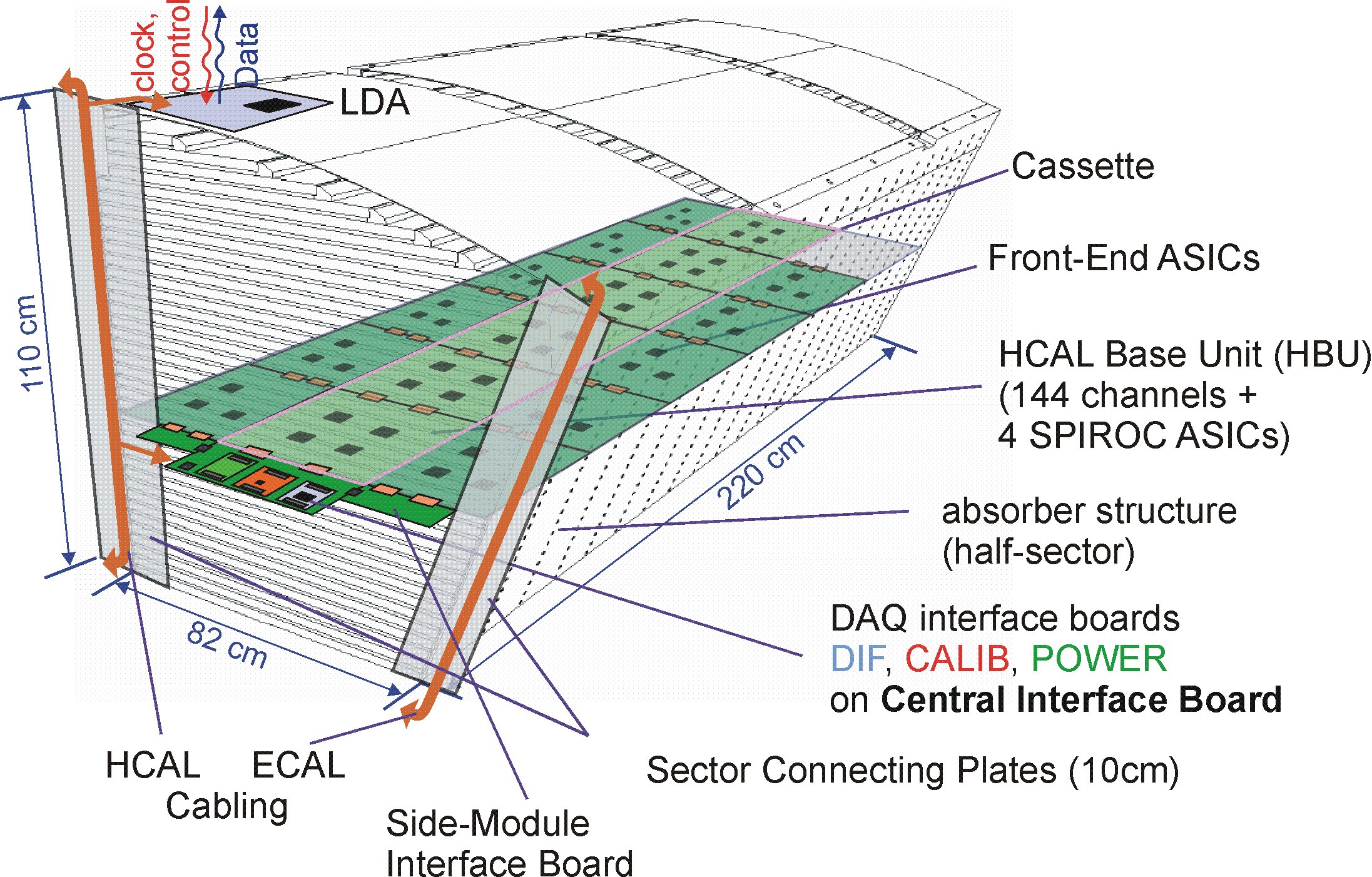}
 	\caption{Structure of $1/16^{th}$ part of the AHCAL's barrel, with only one detector layer shown. The electronics are integrated into the absorber structure.}
 	\label{fig:next-prototype}
 \end{figure}

After the successful data taking with the physical prototype, one project of the CALICE collaboration is to build a technical prototype of the analog hadron colarimeter, which fulfills the requirements of a detector for the International Linear Collider with a realistic mechanical structure and fully integrated readout electronics.
\\
In one design for the International Large Detector, the circular HCAL barrel would be subdivided into 16 sections in the azimuthal direction and into two halfs in the longitudinal direction. One of these halfs is shown in Figure \ref{fig:next-prototype}.  To test mechanical stability, tolerances and deformations, a new integration prototype is currently under development at DESY Hamburg and will cover one layer as shown in Figure \ref{fig:next-prototype}. The typical size of a sector's layer is 1$\times$2.2\,$\textrm{m}^2$.  Every layer consists out of a 2\,cm thick stainless steel absorber plate and so called base units (HCAL base unit: HBU) in cassettes, which are placed in the gap of 0.35\,cm between two absorber plates. Each layer will be equipped with three or four cassettes. In one cassette six HBUs are connected together, forming an electrical layer. 
\\
Each HBU has a size of 36$\times$36\,$\textrm{cm}^2$ and integrates 144 scintillator tiles with the SiPMs, the front-end electronics and the light calibration system. Four SPIROC front-end ASICs are used to read out the analog signal of 144 scintillators on one HBU \cite{prototype}. The HBUs are connected via flexlead to the detector interface, calibration and power modules, which are placed at the end-face of the detector unit for easy access.\\
During assembly, the side-cassettes are assembled first and moved to the layer's sides, and finally the middle cassette is inserted into the structure. The data acquisition (DAQ) interface modules are connected to the middle cassette. For the necessary changes of the layer widths, only the side-cassettes will change from HCAL layer to layer, while the more cost intensive central interface board stays the same for all layers. The room in front of the connecting plates is used for the necessary cabling trees for ECAL and HCAL subdetectors. First beam test of the fully assembled HBUs are currently in progress.

\section{Summary}
Calorimeters with unprecedented granularity have been constructed and tested by the CALICE collaboration for a detector at the future Linear Collider. The high granularity gives the possibility to compare data and hadronic models in GEANT4 simulation via shower properties with high precession. A single cell based weighting technique, based on the hit energy density, was shown to improve the linearity of the reconstructed energy and the energy resolution. As a next step an engineering prototype is under construction, which combines compact electronics with a mechanical stable large scale design.

\end{document}